\documentclass[pdftex,twocolumn,epjc3]{svjour3}  
\RequirePackage[T1]{fontenc}
\smartqed
\RequirePackage{graphicx}
\RequirePackage{flushend}
\RequirePackage[numbers,sort&compress]{natbib}
\RequirePackage[colorlinks,citecolor=blue,urlcolor=blue,linkcolor=blue]{hyperref}

\journalname{Eur. Phys. J. C}

\usepackage{orcidlink}

\usepackage{xcolor}
\usepackage{url}
\usepackage{booktabs}
\usepackage{topcapt}
\usepackage{amsfonts}
\usepackage{nicefrac}
\usepackage{microtype}
\usepackage{amsmath}
\usepackage{multirow}
\usepackage{xspace}
\usepackage{newtxtext,newtxmath}
\usepackage{bbm}
\usepackage{pifont}

\newcommand{\parenthesis}[1]{\left( #1 \right)}

\newcommand{\etal}{et al.\ }

\newcommand{\PW}{\ensuremath{\mathrm{W}\xspace}}
\newcommand{\PZ}{\ensuremath{\mathrm{Z}\xspace}}

\newcommand{\order}[1]{\mathcal{O} \left( #1 \right)}

\newcommand{\jetnet} {{\textsc{JetNet}}\xspace}

\newcommand{\kt}{\ensuremath{k_{\mathrm{T}}}\xspace}
\newcommand{\pt}{\ensuremath{p_{\mathrm{T}}}\xspace}

\newcommand{\PYTHIA} {{\textsc{pythia}}\xspace}

\newcommand{\MADGRAPH} {{\textsc{MADGRAPH}}\xspace}
\newcommand{\MCATNLO} {{\textsc{MCATNLO}}\xspace}
\newcommand{\MGvATNLO}{\MADGRAPH{}5\_a\MCATNLO\xspace}

\newcommand{\etarel}{\ensuremath{\eta^\mathrm{rel}}\xspace}
\newcommand{\phirel}{\ensuremath{\phi^\mathrm{rel}}\xspace}
\newcommand{\ptrel}{\ensuremath{\pt^\mathrm{rel}}\xspace}

\newcommand{\xmark}{\text{\ding{55}}}

\usepackage{lineno}

\begin{document}

\title{Lorentz group equivariant autoencoders}

\author{
    Zichun Hao\thanksref{ucsd}\orcidlink{0000-0002-5624-4907}
    \and
    Raghav Kansal\thanksref{e1,ucsd,fnal}\,\orcidlink{0000-0003-2445-1060}
    \and
    Javier Duarte\thanksref{ucsd}\,\orcidlink{0000-0002-5076-7096}
    \and
    Nadezda Chernyavskaya\thanksref{cern}\,\orcidlink{0000-0002-2264-2229}
}
\thankstext[$\star$]{t1}{
    R.~K. was partially supported by the LHC Physics Center at Fermi National Accelerator Laboratory, managed and operated by Fermi Research Alliance, LLC under Contract No. DE-AC02-07CH11359 with the U.S. Department of Energy (DOE).
    J.~D. and R.~K. were supported by the DOE, Office of Science, Office of High Energy Physics Early Career Research program under Award No. DE-SC0021187, the DOE, Office of Advanced Scientific Computing Research under Award No. DE-SC0021396 (FAIR4HEP), and the NSF HDR Institute for Accelerating AI Algorithms for Data Driven Discovery (A3D3) under Cooperative Agreement OAC-2117997.
    N.~C. was supported by the European Research Council (ERC) under the European Union's Horizon 2020 research and innovation program (Grant Agreement No. 772369).
    This work was performed using the Pacific Research Platform Nautilus HyperCluster supported by NSF awards CNS-1730158, ACI-1540112, ACI-1541349, OAC-1826967, the University of California Office of the President, and the University of California San Diego's California Institute for Telecommunications and Information Technology/Qualcomm Institute.
    Thanks to CENIC for the 100\,Gpbs networks.
}
\thankstext{e1}{e-mail: rkansal@ucsd.edu}

\institute{University of California San Diego, La Jolla, CA, 92093, USA\label{ucsd}
    \and
    Fermi National Accelerator Laboratory, Batavia, IL, 60510, USA\label{fnal}
    \and
    European Organization for Nuclear Research (CERN), 1211, Geneva 23, Switzerland\label{cern}
}

\date{Received: 24 December 2022 / Accepted: 17 May 2023}
\maketitle

\begin{abstract}
    There has been significant work recently in developing machine learning (ML) models in high energy physics (HEP) for tasks such as classification, simulation, and anomaly detection.
    Often these models are adapted from those designed for datasets in computer vision or natural language processing,
    which lack inductive biases suited to HEP data, such as equivariance to its inherent symmetries.
    Such biases have been shown to make models more performant and interpretable, and reduce the amount of training data needed.
    To that end, we develop the Lorentz group autoencoder (LGAE), an autoencoder model equivariant with respect to the proper, orthochronous Lorentz group $\mathrm{SO}^+(3,1)$, with a latent space living in the representations of the group.
    We present our architecture and several experimental results on jets at the LHC and find it outperforms graph and convolutional neural network baseline models on several compression, reconstruction, and anomaly detection metrics.
    We also demonstrate the advantage of such an equivariant model in analyzing the latent space of the autoencoder, which can improve the explainability of potential anomalies discovered by such ML models.
\end{abstract}

\section{Introduction}
\label{sec:intro}

The increasingly large volume of data produced at the LHC and the new era of the High-Luminosity CERN Large Hadron Collider (LHC) poses a significant computational challenge in high energy physics (HEP).
To face this, machine learning (ML) and deep neural networks (DNNs) are becoming powerful and ubiquitous tools for the analysis of particle collisions and their products, such as jets---collimated sprays of particles~\cite{jet_substructure} produced in high energy collisions.

DNNs have been explored extensively for many tasks, such as
classification~\cite{Baldi:2014kfa,Baldi:2014pta,tagging-DNN,ParticleNet},
regression~\cite{Bury:2020ewi,Belayneh:2019vyx},
track reconstruction~\cite{Duarte:2020ngm,track-recons-DNNs,track-recons-IN},
anomaly detection~\cite{AE-anomaly-CNN, Heimel:2018mkt,AE-anomaly,VAE-anomaly-VAE,VAE-anomaly-latent-space, pgae, QUAK},
and simulation~\cite{de_Oliveira_2017, Paganini_2018, Raghav_GAN,sim-VAE-1,sim-VAE-2,CaloGAN}.
\footnote{Interested readers can find comprehensive reviews in Ref.~\cite{Guest_2018,Radovic_2018,Carleo_2019} and a living review in Ref.~\cite{hepmllivingreview}.}
In particular, there has been recent success using networks that incorporate key inductive biases of HEP data, such as infrared and colinear (IRC) safety via energy flow networks~\cite{efn} or graph neural networks (GNNs)~\cite{Konar:2021zdg,Atkinson:2022uzb,Shlomi:2020gdn} and permutation symmetry and sparsity of jet constituents via GNNs~\cite{GNN_Review,ParticleNet,Raghav_GAN}.

Embedding such inductive biases and symmetries into DNNs can not only improve performance, as demonstrated in the references above, but also improve interpretability and reduce the amount of required training data.
Hence, in this paper, we explore another fundamental symmetry of our data: equivariance to Lorentz transformations.
Lorentz symmetry has been successfully exploited recently in HEP for jet classification~\cite{LGN,LorentzNet,Li:2022xfc,Butter_2018}, with competitive and even state-of-the-art (SOTA) results.
We expand this work to the tasks of data compression and anomaly detection by incorporating the Lorentz symmetry into an autoencoder.

Autoencoders learn to encode and decode input data into a learned latent space, and thus have interesting applications in both data compression~\cite{AE-data-compression-1,AE-data-compression-2} and anomaly detection~\cite{AE-anomaly,AE-anomaly-CNN,VAE-anomaly-VAE,pgae,Farina-anomaly,Finke-anomaly,VAE-anomaly-latent-space,QUAK}.
Both tasks are particularly relevant for HEP, the former to cope with the storage and processing of the ever-increasing data collected at the LHC, and the latter for model-independent searches for new physics.
Incorporating Lorentz equivariance into an autoencoder has the potential to not only increase performance in both regards, but also provide a more interpretable latent space and reduce training data requirements.
To this end, in this paper, we develop a Lorentz-group-equivariant autoencoder (LGAE) and explore its performance and interpretability.
We also train alternative architectures, including GNNs and convolutional neural networks (CNNs), with different inherent symmetries and find the LGAE outperforms them on reconstruction and anomaly detection tasks.

The principal results of this work demonstrate (i) that the advantage of incorporating Lorentz equivariance extends beyond whole jet classification to applications with particle-level outputs and (ii) the interpretability of Lorentz-equivariant models.
The key challenges overcome in this work include: (i) training an equivariant autoencoder via particle-to-particle and permutation-invariant set-to-set losses (Section~\ref{sec:experiments}), (ii) defining a jet-level compression scheme for the latent space (Section~\ref{sec:architecture}), and (iii) optimizing the architecture for different tasks, such as reconstruction (Section~\ref{sec:reconstruction}) and anomaly detection (Section~\ref{sec:anomaly}).

This paper is structured as follows.
In Section~\ref{sec:relatedwork}, we discuss existing work, motivating the LGAE.
We present the LGAE architecture in Section~\ref{sec:architecture}, and discuss experimental results on the reconstruction and anomaly detection of high energy jets in Section~\ref{sec:experiments}.
We also demonstrate the interpretability of the model, by analyzing its latent space, and its data efficiency relative to baseline models.
Finally, we conclude in Section~\ref{sec:conclusion}.

\section{Related Work}
\label{sec:relatedwork}
In this section, we briefly review the large body of work on frameworks for equivariant neural networks in Section~\ref{sec:enns}, recent progress in Lorentz-equivariant networks in Section~\ref{sec:lgenn}, and finally, applications of autoencoders in HEP in Section~\ref{sec:aehep}.

\subsection{\label{sec:enns} Equivariant Neural Networks}
A neural network $\mathrm{NN}: V \to W$ is said to be \textit{equivariant} with respect to a group $G$ if
\begin{equation}
    \forall g \in G, v \in V \colon \mathrm{NN} (\rho_V(g) \cdot v) = \rho_W(g) \cdot \mathrm{NN}(v),
\end{equation}
where $\rho_V\colon G \to \mathrm{GL}(V)$ and $\rho_W\colon G \to \mathrm{GL}(W)$ are representations of $G$ in spaces $V$ and $W$ respectively, where $\mathrm{GL}(X)$ is the general linear group of vector space $X$. 
The neural network is said to be \textit{invariant} if $\rho_W$ is a trivial representation, i.e. $\rho_W(g) = \mathbbm{1}_W$ for all $g \in G$.


Equivariance has long been built into a number of successful DNN architectures, such as translation equivariance in CNNs, and permutation equivariance in GNNs~\cite{bronstein2021geometric}.
Recently, equivariance in DNNs has been extended to a broader set of symmetries, such as those corresponding to the
2-dimensional special orthogonal $\mathrm{SO}(2)$~\cite{walters2021trajectory},
the Euclidean $\mathrm{E}(2)$~\cite{E(2)-Equivariant},
the 3-dimensional special orthogonal $\mathrm{SO}(3)$~\cite{SO(3)-Equivariant},
the 3-dimensional Euclidean $\mathrm{E}(3)$~\cite{thomas2018tensor,E3-equivariant} groups,
and arbitrary matrix Lie groups~\cite{Finzi-arbitary-group-equivariance}.

Broadly, equivariance to a group $G$ has been achieved either by extending the translation-equivariant convolutions in CNNs to more general symmetries with appropriately defined learnable filters~\cite{equivariance-kernel-Cohen,equivariance-kernel-Finzi,Cohen-group-equivariance,e3cnn}, or by operating in the Fourier space of $G$, or a combination thereof.
We employ the Fourier space approach, which uses the set of irreducible representations (irreps) of $G$ as the basis for constructing equivariant maps~\cite{equivariance-Fourier-Kondor,equivariance-Fourier-Anderson,E(2)-Equivariant}.


\subsection{\label{sec:lgenn} Lorentz Group Equivariant Neural Networks}

The Lorentz group $\mathrm{O}(3, 1)$ comprises the set of linear transformations between inertial frames with coincident origins.
In this paper, we restrict ourselves to the special orthochronous Lorentz group $\mathrm{SO}^+(3, 1)$, which consists of all Lorentz transformations that preserve the orientation and direction of time.
Lorentz symmetry, or invariance to transformations defined by the Lorentz group, is a fundamental symmetry of the data collected out of high-energy particle collisions.

There have been some recent advances in incorporating this symmetry into NNs.
The Lorentz group network (LGN)~\cite{LGN} was the first DNN architecture developed to be equivariant to the $\mathrm{SO}^+(3, 1)$ group, with an architecture similar to that of a GNN, but operating entirely in Fourier space on objects in irreps of the Lorentz group, and using tensor products between irreps and Clebsch--Gordan decompositions to introduce non-linearities in the network.
More recently, LorentzNet~\cite{LorentzNet,Li:2022xfc} uses a similar GNN framework for equivariance, with additional edge features --- Minkowski inner products between node features --- but restricting itself to only scalar and vector representations of the group.
Both networks have been successful in jet classification, with LorentzNet achieving SOTA results in top quark and quark versus gluon classification, further demonstrating the benefit of incorporating physical inductive biases into network architectures.
In this work, we build on top of the LGN framework to output not only scalars (e.g. jet class probabilities) but encode and reconstruct an input set of particles under the constraint of Lorentz group equivariance in an autoencoder-style architecture.



\begin{figure*}[htb!]
    \centering
    \includegraphics[width=\linewidth]{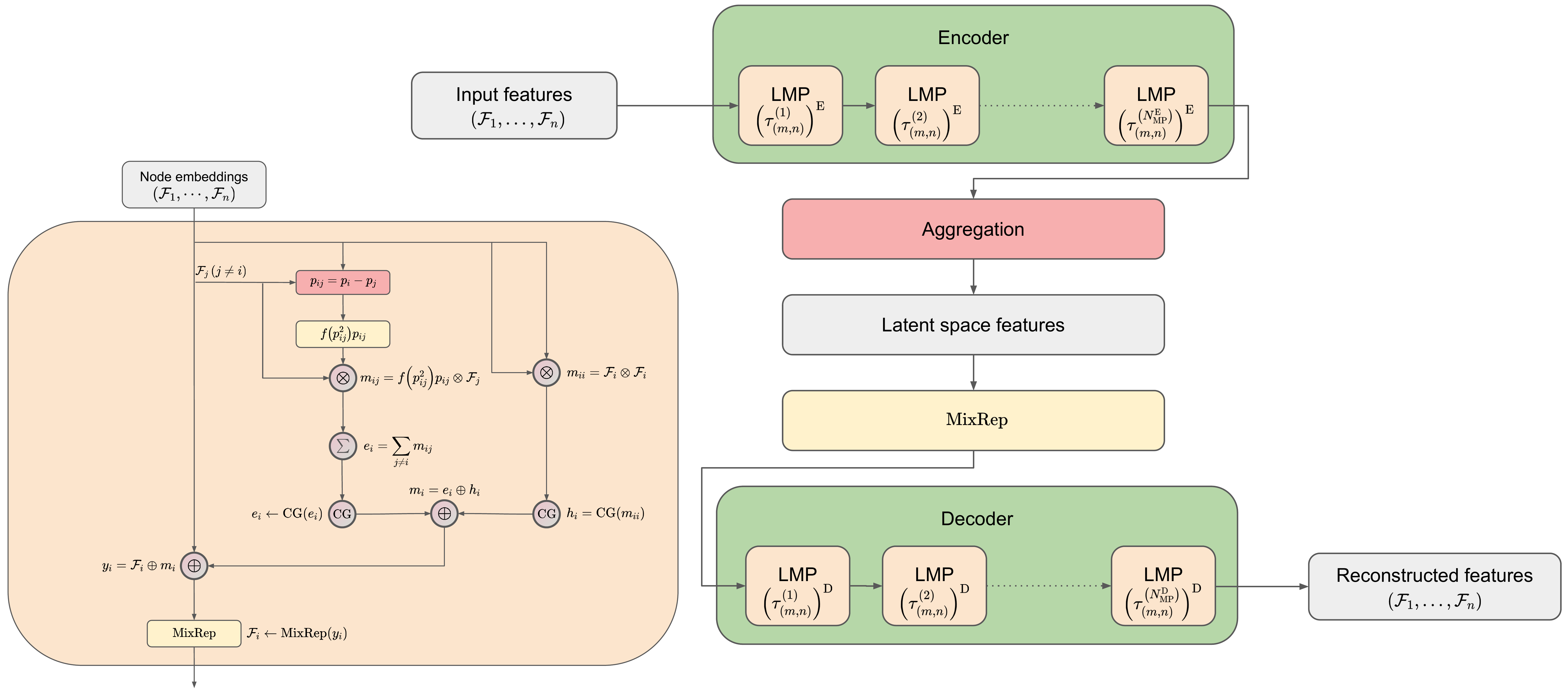}
    \caption{Individual Lorentz group equivariant message passing (LMP) layers are shown on the left, and the LGAE architecture is built out of LMPs on the right. 
    Here, $\mathrm{MixRep}$ denotes the node-level operator that upsamples features in each $(m, n)$ representation space to $\tau_{(m, n)}$ channels; it appears as $W$ in Eq.~(\ref{eq:update}).}
    \label{fig:architecture-full}
\end{figure*}

\subsection{\label{sec:aehep} Autoencoders in HEP}

An autoencoder is an NN architecture comprised of an encoder, which maps the input into a, typically lower dimensional, latent space, and a decoder, which attempts to reconstruct the original input from the latent features.
By using a lower dimensional latent space, an autoencoder can learn a smaller representation of data that captures salient properties~\cite{autoencoders}, which can be valuable in HEP for compressing the significant volumes of data collected at the LHC~\cite{Guglielmo}.

This learned representation can also be exploited for later downstream tasks, such as anomaly detection, where an autoencoder is trained to reconstruct data considered ``background'' to our signal, with the expectation that it will reconstruct the signal poorly relative to the background.
Thus, examining the reconstruction loss of a trained autoencoder may allow the identification of anomalous data\footnote{Another approach directly examines the latent space~\cite{VAE-anomaly-VAE, VAE-anomaly-latent-space}.}.
This can be an advantage in searches for new physics, since instead of having to specify a particular signal hypothesis, a broader search can be performed for data incompatible with the background.
This approach has been successfully demonstrated in Refs.~\cite{Heimel:2018mkt,Farina-anomaly,Cerri:2018anq,Finke-anomaly,Kasieczka:2021xcg,Govorkova:2021utb,Pol:2020weg,Ngairangbam:2021yma,Dillon-lower-dimension}.

Furthermore, there are many possible variations to the general autoencoder framework for alternative tasks~\cite{AE-review-1,AE-review-2}, such as variational autoencoders (VAEs)~\cite{VAE}, which are popular generative models.
To our knowledge, while there have been some recent efforts at GNN-based autoencoder models~\cite{pgae,Atkinson:2021nlt}, Lorentz equivariance has not yet been explored.
In this work, we focus on data compression and anomaly detection but note that our model can be extended to further applications.

\section{LGAE architecture}
\label{sec:architecture}
The LGAE is built out of Lorentz group-equivariant message passing (LMP) layers, which are identical to individual layers in the LGN~\cite{LGN}.
We reinterpret them in the framework of message-passing neural networks~\cite{Gilmer-GNN}, to highlight the connection to GNNs, and define them in Sec.~\ref{sec:lmp}.
We then describe the encoder and decoder networks in Secs.~\ref{sec:lgaee} and \ref{sec:lgaed}, respectively.
The LMP layers and LGAE architecture are depicted in Fig.~\ref{fig:architecture-full}.
We provide the LGAE code, written in Python using the \textsc{PyTorch} ML framework~\cite{pytorch} in Ref.~\cite{LGAE_code}.

\subsection{LMP \label{sec:lmp}}

LMP layers take as inputs fully-connected graphs with nodes representing particles and the Minkowski distance between respective node 4-vectors as edge features. 
Each node $\mathcal{F}_i$ is defined by its features, all transforming under a corresponding irrep of the Lorentz group in the canonical basis~\cite{gelfand1963}, including at least one 4-vector (transforming under the $(1/2, 1/2)$ representation) representing its 4-momentum.
As in Ref~\cite{LGN}, we denote the number of features in each node transforming under the $(m, n)$ irrep as $\tau_{(m,n)}$, referred to as the multiplicity of the $(m,n)$ representation. 

The $(t+1)$-th MP layer operation consists of message-passing between each pair of nodes, with a message $m_{i j}^{(t)}$ to node $i$ from node $j$ (where $j \neq i$) and a self-interaction term $m_{ii}$ defined as
\begin{align} \label{eq:msg}
        m_{i j}^{(t)} &= f\left( \left(p_{ij}^{(t)}\right)^2 \right) p_{ij}^{(t)} \otimes \mathcal{F}_j^{(t)} \\ 
        m_{i i}^{(t)} &= \mathcal{F}_i^{(t)} \otimes \mathcal{F}_i^{(t)}
\end{align}
where $\mathcal{F}_{i}^{(t)}$ are the node features of node $i$ before the $(t+1)$-th layer, $p_{ij} = p_i - p_j$ is the difference between node four-vectors, $p_{ij}^2$ is the squared Minkowski norm of $p_{i j}$, and $f$ is a learnable, differentiable function acting on Lorentz scalars. 
A Clebsch--Gordan (CG) decomposition, which reduces the features to direct sums of irreps of $\mathrm{SO}^+(3,1)$, is performed on both terms before concatenating them to produce the message $m_i$ for node $i$:
\begin{equation}
    m_i^{(t)} = 
    \mathrm{CG}\left[
    m_{i i}^{(t)}
    \right]
    \oplus
    \mathrm{CG}\left[
    \sum_{j\neq i} m_{i j}^{(t)}
    \right], 
\end{equation}
where the summation over the destination node $j$ ensures permutation symmetry because it treats all other nodes equally. 

Finally, this aggregated message is used to update each node's features, such that
\begin{equation} \label{eq:update}
    \mathcal{F}_i^{(t+1)} = W^{(t+1)} \left( \mathcal{F}_i^{(t)} \oplus m_i^{(t)} \right)
\end{equation}
for all $i \in \{1, \ldots, N_\mathrm{particle}\}$, where $W^{(t+1)}$ is a learnable node-wise operator which acts as separate fully-connected linear layers $W^{(t+1)}_{(m, n)}$ on the set of components living within each separate $(m, n)$ representation space, outputting a chosen $\tau_{(m,n)}^{(t+1)}$ number of components per representation.
In practice, we then truncate the irreps to a maximum dimension to make computations more tractable.

\subsection{Encoder \label{sec:lgaee}}
The encoder takes as input an $N$-particle cloud, where each particle is each associated with a 4-momentum vector and an arbitrary number of scalars representing physical features such as mass, charge, and spin. 
Each isotypic component is initially transformed to a chosen multiplicity of $\parenthesis{\tau_{(m, n)}^{(0)}}_\mathrm{E}$ via a node-wise operator $W^{(0)}$ identical conceptually to $W^{(t+1)}$ in Eq.~(\ref{eq:update}). 
The resultant graph is then processed through $N_{\mathrm{MP}}^\mathrm{E}$ LMP layers, specified by a sequence of multiplicities $\left\{ \parenthesis{\tau_{(m, n)}^{(t)}}_\mathrm{E} \right\}_{t=1}^{N_{\mathrm{MP}}^\mathrm{E}}$, where $\parenthesis{\tau_{(m, n)}^{(t)}}_\mathrm{E}$ is the multiplicity of the $(m, n)$ representation at the $t$-th layer. 
Weights are shared across the nodes in a layer to ensure permutation equivariance.

After the final MP layer, node features are aggregated to the latent space by a component-wise minimum (min), maximum (max), or mean. 
The min and max operations are performed on the respective Lorentz invariants.
We also find, empirically, interesting performance by simply concatenating isotypic components across each particle and linearly ``mixing" them via a learned matrix as in Eq.~(\ref{eq:update}). 
Crucially, unlike in Eq.~(\ref{eq:update}), where this operation only happens per particle, the concatenation across the particles imposes an ordering and, hence, breaks the permutation symmetry.



\subsection{Decoder \label{sec:lgaed}}
The decoder recovers the $N$-particle cloud by acting on the latent space with $N$ independent, learned linear operators, which again mix components living in the same representations.
This cloud passes through $N_{\mathrm{MP}}^\mathrm{D}$ LMP layers, specified by a sequence of multiplicities $\left\{ \parenthesis{\tau_{(m, n)}^{(t)}}_\mathrm{D} \right\}_{t=1}^{N_{\mathrm{MP}}^\mathrm{D}}$, where $\parenthesis{\tau_{(m, n)}^{(t)}}_\mathrm{D}$ is the multiplicity of the $(m, n)$ representation at the $t$-th LMP layer.
After the LMP layers, node features are mixed back to the input representation space $\parenthesis{D^{(0,0)}}^{\oplus \tau_{(0,0)}^{(0)}} \oplus D^{(1/2, 1/2)}$ by applying a linear mixing layer and then truncating other isotypic components. 

\section{Experiments}
\label{sec:experiments}

We experiment with and evaluate the performance of the LGAE and baseline models on reconstruction and anomaly detection for simulated high-momentum jets.
We describe the dataset in Sec.~\ref{sec:dataset}, the different models we consider in Sec.~\ref{sec:models}, the reconstruction and anomaly detection results in Sec.s~\ref{sec:reconstruction} and~\ref{sec:anomaly} respectively, an interpretation of the LGAE latent space in Sec.~\ref{sec:latent-analysis}, and finally experiments of the data efficiency of the different models in Sec.~\ref{sec:data-efficiency}.

\subsection{Dataset}
\label{sec:dataset}
The model is trained to reconstruct 30-particle high transverse momentum jets from the \jetnet~\cite{jetnetdataset} dataset, obtained using the associated library~\cite{jetnetlibrary}, zero-padding jets with fewer than 30, produced from gluons and light quarks.
These are collectively referred to as quantum chromodynamics (QCD) jets.

Jets in \jetnet are first produced at leading-order using \MGvATNLO~\cite{alwall_madgraph5} and decayed and showered with \PYTHIA8.2~\cite{pythia}.
They are then discretized and smeared to take detector spatial and energy resolution respectively into account, with simulated tracking inefficiencies---emulating the effects of the CMS and ATLAS trackers and calorimeters---and finally clustered using the anti-$\kt$~\cite{antikt} algorithm with distance parameter $R=0.8$.
Further details on the generation and reconstruction process are available in Ref.~\cite{Raghav_GAN}.
The exact smearing parameters and calorimeter granularities used are reported in Table~2 of Ref.~\cite{Coleman:2017fiq} and correspond to the ``CMS-like'' scenario.

\begin{table*}[t!]
    \centering
    \caption{Summary of the relevant symmetries respected by each model discussed in Sec.~\ref{sec:experiments}.}
    \label{tab:model-symmetry}
    \begin{tabular*}{\linewidth}{@{\extracolsep{\fill}} llllll }
        \toprule
        Model                  & Aggregation    & Name                    & Lorentz symmetry        & Permutation symmetry  & Translation symmetry  \\ \midrule
        \multirow{2}{*}{LGAE}  &
        Min-Max                & LGAE-Min-Max   & \checkmark (equivariance) & \checkmark (invariance)  & \checkmark (equivariance)                           \\
                               & Mix            & LGAE-Mix                & \checkmark (equivariance) & \xmark   & \checkmark (equivariance)               \\[2mm]
        \multirow{2}{*}{GNNAE} &
        Jet-level              & GNNAE-JL       & \xmark                  & \checkmark (invariance) & \checkmark (equivariance)                          \\
                               & Particle-level & GNNAE-PL                & \xmark                  & \checkmark (equivariance) & \checkmark (equivariance) \\[2mm]
        CNNAE & & CNNAE & \xmark  & \xmark & \checkmark (equivariance) \\
        \bottomrule
    \end{tabular*}
\end{table*}

We represent the jets as a point cloud of particles, termed a ``particle cloud``, with the respective 3-momenta, in absolute coordinates, as particle features.
In the processing step, each 3-momentum is converted to a 4-momentum: $p^\mu = (|\mathbf{p}|, \mathbf{p})$, where we consider the mass of each particle to be negligible.
We use a $60\%/20\%/20\%$ training/testing/validation splitting for the total 177,000 jets.
For evaluating performance in anomaly detection, we consider jets from \jetnet produced by top quarks, $W$ bosons, and $Z$ bosons as our anomalous signals.

Finally, we note here that the detector and reconstruction effects in \jetnet, and indeed in real data collected at the LHC, break the Lorentz symmetry; hence, Lorentz equivariance is generally an \textit{approximate} rather than an exact symmetry of HEP data.
We assume henceforth that the magnitude of the symmetry breaking is small enough that imposing exact Lorentz equivariance in the LGAE is still advantageous---and the high performance of the LGAE and classification models such as LorentzNet support this assumption.
Nevertheless, important studies in future work may include quantifying this symmetry breaking and considering approximate, as well as exact, symmetries in neural networks.

\subsection{Models}
\label{sec:models}
LGAE model results are presented using both the min-max (LGAE-Min-Max) and ``mix'' (LGAE-Mix) aggregation schemes for the latent space, which consists of varying numbers of complex Lorentz vectors --- corresponding to different compression rates.
We compare the LGAE to baseline GNN and CNN autoencoder models, referred to as ``GNNAE'' and ``CNNAE'' respectively. 

The GNNAE model is composed of fully-connected MPNNs adapted from Ref.~\cite{Raghav_GAN}.
We experiment with two types of encodings: (1) particle-level (GNNAE-PL), as in the PGAE~\cite{pgae} model, which compresses the features per node in the graph but retains the graph structure in the latent space, and (2) jet-level (GNNAE-JL), which averages the features across each node to form the latent space, as in the LGAE.
Particle-level encodings produce better performance overall for the GNNAE, but the jet-level provides a more fair comparison with the LGAE, which uses jet-level encoding to achieve a high level of compression of the features.

For the CNNAE, which is adapted from Ref.~\cite{farina_cnnae}, the relative coordinates of each input jets' particle constituents are first discretized into a $40 \times 40$ grid.
The particles are then represented as pixels in an image, with intensities corresponding to $\ptrel$.
Multiple particles per jet may correspond to the same pixel, in which case their \ptrel's are summed.
The CNNAE has neither Lorentz nor permutation symmetry, however, it does have in-built translation equivariance in $\eta-\phi$ space.

Hyperparameter and training details for all models can be found in~\ref{app:hyperparams} and~\ref{app:training} respectively, and a summary of the relevant symmetries respected by each model is provided in Table~\ref{tab:model-symmetry}.
The LGAE models are verified to be equivariant to Lorentz boosts and rotations up to numerical error, with details provided in~\ref{app:equivariancetests}.


\subsection{Reconstruction}
\label{sec:reconstruction}
\begin{figure*}[htpb]
    \centering
    \includegraphics[width=\linewidth]{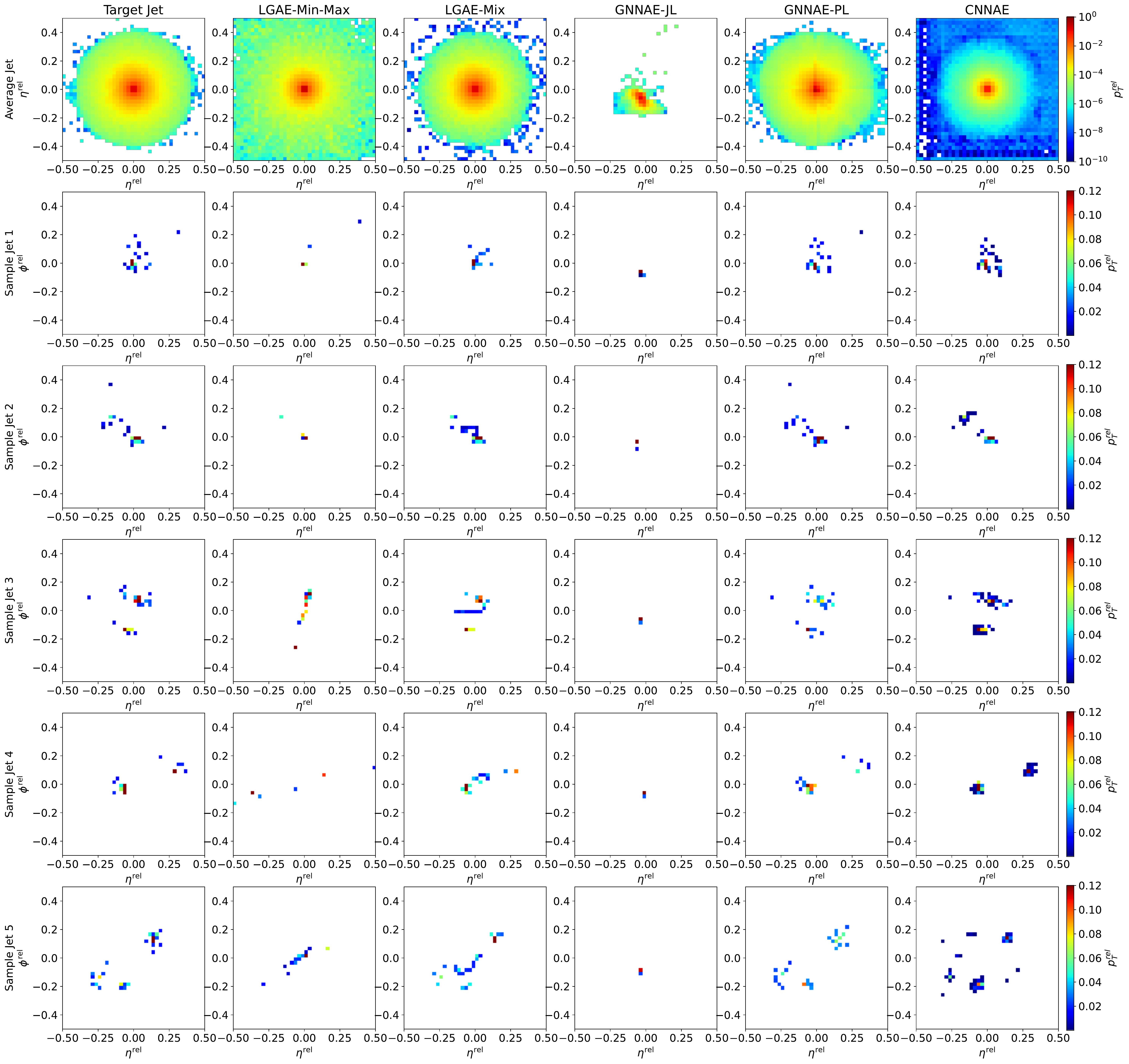}
    \caption{Jet image reconstructions by
        LGAE-Min-Max ($\tau_{(1/2, 1/2)}=4$, $56.67\%$ compression),
        LGAE-Mix ($\tau_{(1/2, 1/2)}=9$, $61.67\%$ compression),
        GNNAE-JL ($\dim(L) = 55$, $61.11\%$ compression), 
        GNNAE-PL ($\dim(L) = 2\times 30$, $66.67\%$ compression), and CNNAE ($\dim(L) = 55$, $61.11\%$ compression).
    }
    \label{fig:recons-jet-imgs}
\end{figure*}

We evaluate the performance of the LGAE, GNNAE, and CNNAE models, with the different aggregation schemes discussed, on the reconstruction of the particle and jet features of QCD jets.  
We consider relative transverse momentum
$\ptrel = \pt^\mathrm{particle}/\pt^\mathrm{jet}$ and relative angular coordinates
$\etarel =\eta^\mathrm{particle} - \eta^\mathrm{jet}$ and
$\phirel =\phi^\mathrm{particle} - \phi^\mathrm{jet} \pmod{2\pi}$
as each particle's features, and total jet mass, \pt and $\eta$ as jet features. 
We define the compression rate as the ratio between the total dimension of the latent space and the number of features in the input space: $30\ \mathrm{particles} \times 3\ \mathrm{features\ per\ particle} = 90$.

\begin{figure*}[t]
    \centering
    \includegraphics[width=\linewidth]{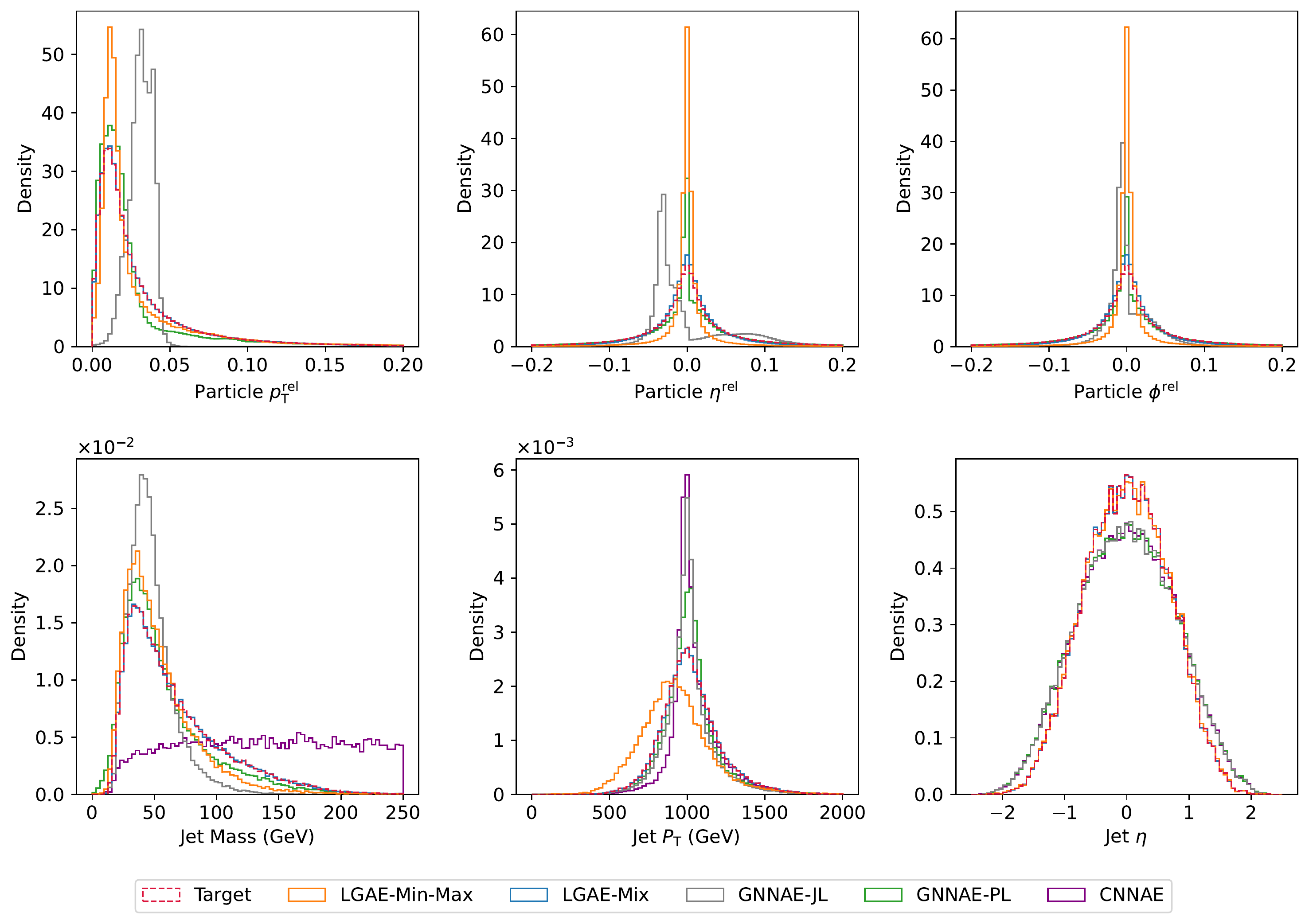}
    \caption{
        \textbf{Top}:
        particle momenta $(\pt^\mathrm{rel}, \eta^\mathrm{rel}, \phi^\mathrm{rel})$ reconstruction by
        LGAE-Min-Max ($\tau_{(1/2, 1/2)}=4$, resulting in $56.67\%$ compression) and
        and LGAE-Mix ($\tau_{(1/2, 1/2)}=9$, resulting in $61.67\%$ compression),
        and GNNAE-JL ($\dim(L) = 55$, resulting in $61.11\%$ compression) and
        GNNAE-PL ($\dim(L) = 2\times 30$, resulting in $66.67\%$ compression). The reconstructions by the CNNAE are not included due to the discrete values of $\etarel$ and $\phirel$, as discussed in the text. 
        \textbf{Bottom}:
        jet feature $(M, \pt, \eta)$ reconstruction by the four models.
        For the jet feature reconstruction by the GNNAEs, the particle features in relative coordinates were transformed back to absolute coordinates before plotting.
        The jet $\phi$ is not shown because it follows a uniform distribution in $(-\pi, \pi]$ and is reconstructed well.
    }
    \label{fig:recons-hist}
\end{figure*}

Figure~\ref{fig:recons-jet-imgs} shows random samples of jets, represented as discrete images in the angular-coordinate plane, reconstructed by the models with similar levels of compression in comparison to the true jets.
Figure~\ref{fig:recons-hist} shows histograms of the reconstructed features compared to the true distributions.
The differences between the two distributions are quantified in Table~\ref{tab:recons-particle} by calculating the median and interquartile ranges (IQR) of the relative errors between the reconstructed and true features.
To calculate the relative errors of particle features for the permutation invariant LGAE and GNNAE models, particles are matched between the input and output clouds using the Jonker–Volgenant algorithm~\cite{JonkerVolgenant,scipy} based on the L2 distance between particle features.
Due to the discretization of the inputs to the CNNAE, reconstructing individual particle features is not possible; instead, only jet features are shown.\footnote{These are calculated by summing each pixel's momentum ``4-vector'' --- using the center of the pixel as angular coordinates and intensity as the \ptrel.}

We can observe visually in Figure~\ref{fig:recons-jet-imgs} that out of the two permutation invariant models, while neither is able to reconstruct the jet substructure perfectly,
the LGAE-Min-Max outperforms the GNNAE-JL.
Perhaps surprisingly, the permutation-symmetry-breaking mix aggregation scheme improves the LGAE in this regard.
Both visually in Figure~\ref{fig:recons-hist} and quantitatively from Tables~\ref{tab:recons-particle} and~\ref{tab:recons-jet}, we conclude that the LGAE-Mix has the best performance overall, significantly outperforming the GNNAE and CNNAE models at similar compression rates. 
The LGAE-Min-Max model outperforms the GNNAE-JL in reconstructing all features and the GNNAE-PL in all but the IQR of the particle angular coordinates.

\begin{table*}[htpb!]
    \centering
    \caption{Median and IQR of relative errors in particle feature reconstruction of selected LGAE and GNNAE models.
        In each column, the best-performing latent space per model is italicized, and the best model overall is highlighted in bold.
    }
    \label{tab:recons-particle}
    \begin{tabular*}{\linewidth}{@{\extracolsep{\fill}} lllcccccc }  \toprule
        \multirow{2}{*}{Model}
        & \multirow{2}{*}{Aggregation}
        & \multirow{2}{*}{Latent space}
        & \multicolumn{2}{c}{Particle \ptrel}
        & \multicolumn{2}{c}{Particle \etarel}
        & \multicolumn{2}{c}{Particle \phirel}
        \\ \cline{4-9}
        &&& Median & IQR & Median & IQR & Median & IQR
        \\ \hline
        \multirow{4}{*}{LGAE} & \multirow{2}{*}{Min-max}
        & $\tau_{(1/2, 1/2)} = 4$ ($56.67\%$)
        & $\mathit{0.006}$ & $\mathit{0.562}$
        & $\mathit{0.002}$ & ${1.8}$
        & ${0.003}$ & ${1.8}$                              \\ &
        & $\tau_{(1/2, 1/2)} = 7$ ($96.67\%$)
        & $0.002$ & $0.640$
        & $-0.627$ & $\mathit{1.7}$
        & $\mathbf{< 10^{-3}}$ & $\mathit{1.7}$                       \\[1mm]
        & \multirow{2}{*}{Mix}
        & $\tau_{(1/2, 1/2)} = 9$ ($61.67\%$)
        & $\mathbf{< 10^{-3}}$ & $0.011$
        & $\mathbf{< 10^{-3}}$ & $0.452$
        & $\mathbf{< 10^{-3}}$ & ${0.451}$
        \\
        & & $\tau_{(1/2, 1/2)} = 13$ ($88.33\%$)
        & $\mathbf{< 10^{-3}}$ & $\mathbf{0.001}$
        & $\mathbf{< 10^{-3}}$ & $\mathbf{0.022}$
        & $\mathbf{< 10^{-3}}$ & $\mathbf{0.022}$                     \\[2mm]
        \multirow{4}{*}{GNNAE}
    & \multirow{2}{*}{Jet-level}
    & $\dim(L) = 45$ ($50.00\%$)
    & $-0.983$ & $3.8$
    & $\mathit{0.363}$ & $\mathit{3.1}$
    & $\mathit{0.146}$ & $\mathit{2.1}$                       \\
    & & $\dim(L) = 90$ ($100.00\%$)
    & $\mathit{-0.627}$ & $\mathit{3.5}$
    & $4.4$ & ${14.7}$
    & $\mathit{0.146}$ & ${2.6}$                              \\[1mm]
    & \multirow{2}{*}{Particle-level}
    & $\dim(L) = 2 \times 30$ ($66.67\%$)
    & $-0.053$ & $0.906$
    & $\mathit{0.009}$ & ${0.191}$
    & ${0.013}$ & $\mathit{0.139}$                     \\
    & & $\dim(L) = 3 \times 30$ ($100.00\%$)
    & $\mathit{-0.040}$ & $\mathit{0.892}$
    & ${-0.037}$ & $\mathit{0.177}$
    & $\mathit{0.005}$ & $0.243$                              \\
        \bottomrule
    \end{tabular*}
\end{table*}

\begin{table*}[ht]
    \centering
    \caption{Median and IQR of relative errors in jet feature
        reconstruction by selected LGAE and GNNAE models, along with the CNNAE model. 
        In each column, the best performing latent space per model is italicised, and the best model overall is highlighted in bold.
    }
    \label{tab:recons-jet}
    \begin{tabular}{lllcccccccc}  \toprule
        \multirow{2}{*}{Model}
         & \multirow{2}{*}{Aggregation}
         & \multirow{2}{*}{Latent space}
         & \multicolumn{2}{c}{Jet mass}
         & \multicolumn{2}{c}{Jet $\pt$}
         & \multicolumn{2}{c}{Jet $\eta$}
         & \multicolumn{2}{c}{Jet $\phi$}
        \\ \cline{4-11}
         & &
         & Median & IQR
         & Median & IQR
         & Median & IQR
         & Median & IQR
        \\ \hline
        \multirow{4}{*}{LGAE}
         & \multirow{2}{*}{Min-max}
         & $\tau_{(1/2,1/2)} = 4$ ($56.67\%$)
         & $\mathit{0.096}$ & $\mathit{0.134}$
         & $\mathit{0.097}$ & $\mathit{0.109}$
         & $\mathbf{< 10^{-3}}$ & $\mathit{0.004}$
         & $\mathbf{< 10^{-3}}$ & $\mathit{0.002}$
        \\
         & & $\tau_{(1/2,1/2)} = 7$ ($96.67\%$)
         & ${-0.139}$ & ${0.287}$
         & ${-0.221}$ & ${0.609}$
         & $\mathbf{< 10^{-3}}$ & ${0.021}$
         & $\mathbf{< 10^{-3}}$ & ${0.007}$
        \\[1mm]
         & \multirow{2}{*}{Mix}
         & $\tau_{(1/2,1/2)} = 9$ ($61.67\%$)
         & $\mathbf{< 10^{-3}}$ & $\mathbf{0.003}$
         & $\mathbf{< 10^{-3}}$ & $\mathbf{< 10^{-3}}$
         & $\mathbf{< 10^{-3}}$ 
         & $\mathbf{< 10^{-3}}$
         & $\mathbf{< 10^{-3}}$ & 
         $\mathbf{< 10^{-3}}$
        \\
         & & $\tau_{(1/2,1/2)} = 13$ ($88.33\%$)
         & $\mathbf{< 10^{-3}}$ & $\mathbf{0.003}$
         & $\mathbf{< 10^{-3}}$ & $\mathbf{< 10^{-3}}$
         & $\mathbf{< 10^{-3}}$ 
         & $\mathbf{< 10^{-3}}$
         & $\mathbf{< 10^{-3}}$ & 
         $\mathbf{< 10^{-3}}$
        \\[2mm]
        \multirow{4}{*}{GNNAE}
         & \multirow{2}{*}{Jet-level}
         & $\dim(L) = 45$ ($50.00\%$)
         & ${0.326}$ & $\mathit{0.667}$
         & $\mathit{0.030}$ & $\mathit{0.088}$
         & $\mathit{0.005}$ & $\mathit{0.040}$
         & $\mathit{0.001}$ & ${0.021}$
        \\
         & & $\dim(L) = 90$ ($100.00\%$)
         & ${3.7}$ & $2.6$
         & $\mathit{0.030}$ & ${0.089}$
         & ${0.292}$ & ${0.433}$
         & ${0.006}$ & ${0.021}$
        \\[1mm]
         & \multirow{2}{*}{Particle-level}
         & $\dim(L) = 2 \times 30$ ($66.67\%$)
         & $\mathit{0.277}$ & ${0.299}$
         & $\mathit{0.037}$ & ${0.110}$
         & ${0.002}$ & $\mathit{0.010}$
         & $-0.001$ & ${0.005}$
        \\
         & & $\dim(L) = 3 \times 30$ ($100.00\%$)
         & ${0.339}$ & $\mathit{0.244}$
         & ${0.050}$ & $\mathit{0.094}$
         & $\mathit{-0.001}$ & ${0.011}$
         & $\mathbf{<10^{-3}}$ & ${0.005}$
        \\[2mm]
        CNNAE 
        & Linear layer
        & $\dim(L) = 55$ ($61.67\%$) 
        & $-0.030$
        & $0.042$ 
        & $-0.021$
        & $0.017$
        & $\mathbf{< 10^{-3}}$
        & $0.017$
        & $\mathbf{<10^{-3}}$
        & $0.003$ \\
        \bottomrule
    \end{tabular}
\end{table*}

\subsection{Anomaly detection}
\label{sec:anomaly}
\begin{figure*}
    \centering
    \includegraphics[width=\linewidth]{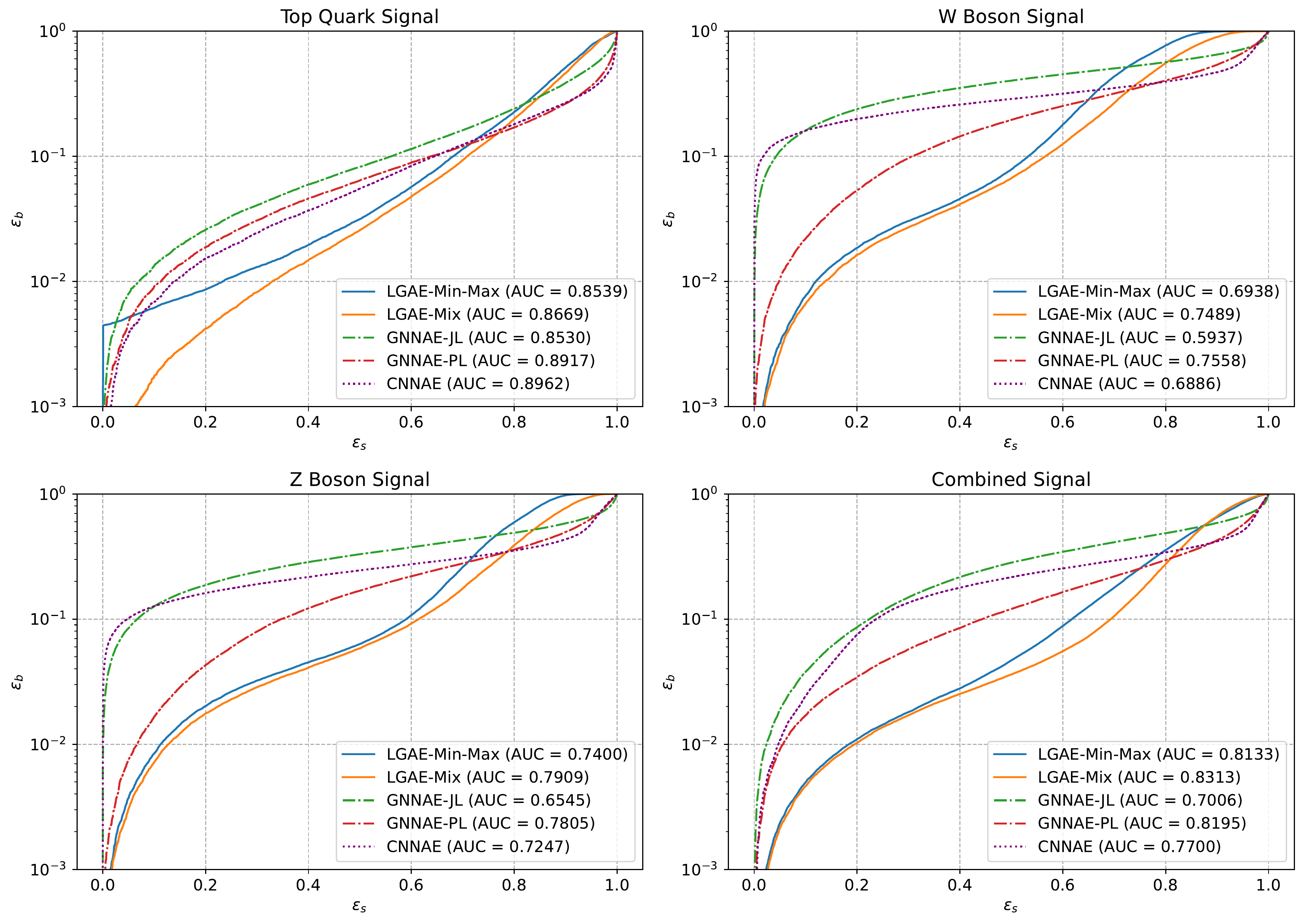}
    \caption{
        Anomaly detections for
        the top quark signal (upper left),
        $W$ boson signal (upper right),
        $Z$ boson signal (lower left),
        and the combined signal (lower right)
        by the selected
        LGAE-Min-Max ($\tau_{(1/2, 1/2)} = 7$),
        LGAE-Mix ($\tau_{(1/2, 1/2)}=2$),
        GNNAE-JL ($\dim(L) = 30$),
        GNNAE-PL ($\dim(L) = 2 \times 30$),
        and CNNAE ($\dim(L) = 55$)
        models.
    }
    \label{fig:roc-each}
\end{figure*}

We test the performance of all models as unsupervised anomaly detection algorithms by pre-training them solely on QCD and then using the reconstruction error for the QCD and new signal jets as the discriminating variable.
We consider top quark, $\PW$ boson, and $\PZ$ boson jets as potential signals and QCD as the ``background''.
We test the Chamfer distance, energy mover's distance~\cite{emd} --- the earth mover's distance applied to particle clouds, and MSE between input and output jets as reconstruction errors, and find the Chamfer distance most performant for all graph-based models. 
For the CNNAE, we use the MSE between the input and reconstructed image as the anomaly score. 

Receiver operating characteristic (ROC) curves showing the signal efficiencies ($\varepsilon_s$) versus background efficiencies ($\varepsilon_b$) for individual and combined signals are shown in Fig.~\ref{fig:roc-each},\footnote{Discontinuities in the top quark and combined signal LGAE-Min-Max ROCs indicate that at background efficiencies of $\lessapprox 5\times10^{-3}$, there are no signal events remaining in the validation dataset.} and $\varepsilon_s$ values at particular background efficiencies are given in Table~\ref{tab:ad}.
We see that in general the permutation equivariant LGAE and GNNAE models outperform the CNNAE, strengthening the case for considering equivariance in neural networks.
Furthermore, LGAE models have significantly higher signal efficiencies than GNNAEs and CNNAEs for all signals when rejecting $>90\%$ of the background (which is the minimum level we typically require in HEP), and LGAE-Mix consistently performs better than LGAE-Min-Max. 


\begin{table*}[htpb]
    \centering
    \caption{
        Anomaly detection metrics by a selected LGAE and GNNAE models, along with the CNNAE model. 
        In each column, the best performing latent space per model is italicized, and the best model overall is highlighted in bold.
    }
    \label{tab:ad}
    \begin{tabular*}{\linewidth}{@{\extracolsep{\fill}} llllccc } \toprule
        \multirow{2}{*}{Model}       &
        \multirow{2}{*}{Aggregation} & \multirow{2}{*}{Latent space}       & \multirow{2}{*}{AUC}                 &
        \multicolumn{3}{c}{$\varepsilon_s$ at given $\varepsilon_b$}
        \\ \cline{5-7}
        & & & & $\varepsilon_s (10^{-1})$ & $\varepsilon_s (10^{-2})$ & $\varepsilon_s (10^{-3})$ \\ \hline
        \multirow{6}{*}{LGAE}
        & \multirow{3}{*}{Min-Max}
        & $\tau_{(1/2,1/2)} = 2$ ($30.00\%$)  & $0.7253$ & $0.5706$ & $0.1130$ & $\mathit{0.0011}$                                     \\
                                     &                                     & $\tau_{(1/2,1/2)} = 4$ ($56.67\%$)   & $0.7627$          & $0.5832$                  & $\mathit{0.1305}$         & $0.0007$                  \\
                                     &                                     & $\tau_{(1/2,1/2)} = 7$ ($96.67\%$)   & $\mathit{0.7673}$ & $\mathit{0.5932}$         & $0.0820$                  & $0.0009$                  \\[1mm]
                                     & \multirow{3}{*}{Mix}
                                     & $\tau_{(1/2,1/2)} = 2$ ($15.00\%$)  & $\mathit{0.8023}$                    & $0.6178$          & $\mathbf{0.1662}$         & $\mathbf{0.0250}$                                     \\
                                     &                                     & $\tau_{(1/2,1/2)} = 4$ ($28.33\%$)   & $\mathit{0.8023}$ & $0.6257$                  & $0.1592$                  & $0.0229$                  \\
                                     &                                     & $\tau_{(1/2,1/2)} = 7$ ($48.33\%$)   & $0.7967$          & $\mathbf{0.6290}$         & $0.1562$                  & $0.0225$                  \\[2mm]
        \multirow{5}{*}{GNNAE}
                                     & \multirow{3}{*}{JL}
                                     & $\dim(L) = 10$ ($11.11\%$)          & $0.5891$                             & $0.1576$          & $0.0161$                  & $\mathit{0.0014}$                                     \\
                                     &                                     & $\dim(L) = 40$ ($44.44\%$)           & $0.6636$          & $\mathit{0.2293}$         & $\mathit{0.0262}$         & $0.0013$                  \\
                                     &                                     & $\dim(L) = 80$ ($88.89\%$)           & $\mathit{0.7006}$ & $0.2240$                  & $0.0239$                  & $0.0010$                  \\[1mm]
                                     & \multirow{2}{*}{PL}
                                     & $\dim(L) = 2 \times 30$ ($66.67\%$) & $\mathbf{0.8195}$                    & $\mathit{0.4435}$ & $0.0564$                  & $0.0042$                                              \\
                                     &                                     & $\dim(L) = 3 \times 30$ ($100.00\%$) & $0.8095$          & $0.4306$                  & $\mathit{0.0762}$         & $\mathit{0.0044}$         \\[2mm]
       CNNAE & linear layer
       & $\dim(L) = 55$ ($61.67\%$)
       & $0.7700$
       & $0.2473$
       & $0.0469$
       & $0.0053$ \\
        \bottomrule
    \end{tabular*}
\end{table*}

\subsection{Latent space interpretation}
\label{sec:latent-analysis}
\begin{figure*}[ht]
    \centering
    \includegraphics[width=\linewidth]{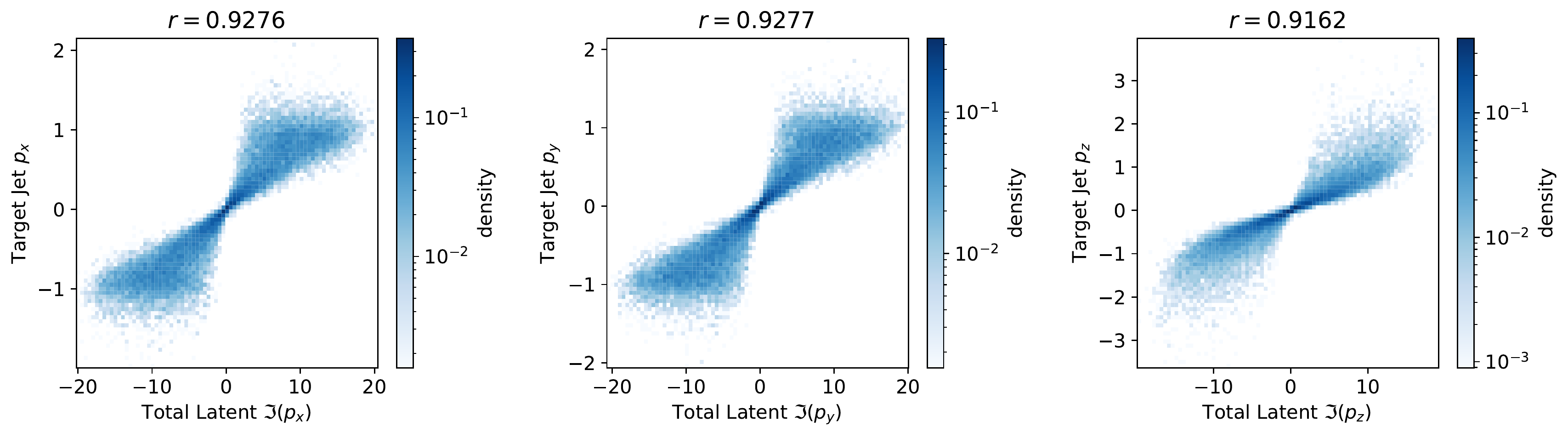}
    \caption{The correlations between the total momentum of the imaginary components in the $\tau_{(1/2, 1/2)} = 2$ LGAE-Mix model and the target jet momenta.
        The Pearson correlation coefficient $r$ is listed above.
    }
    \label{fig:correlations}
\end{figure*}

\begin{figure*}[ht]
    \centering
    \includegraphics[width=\linewidth]{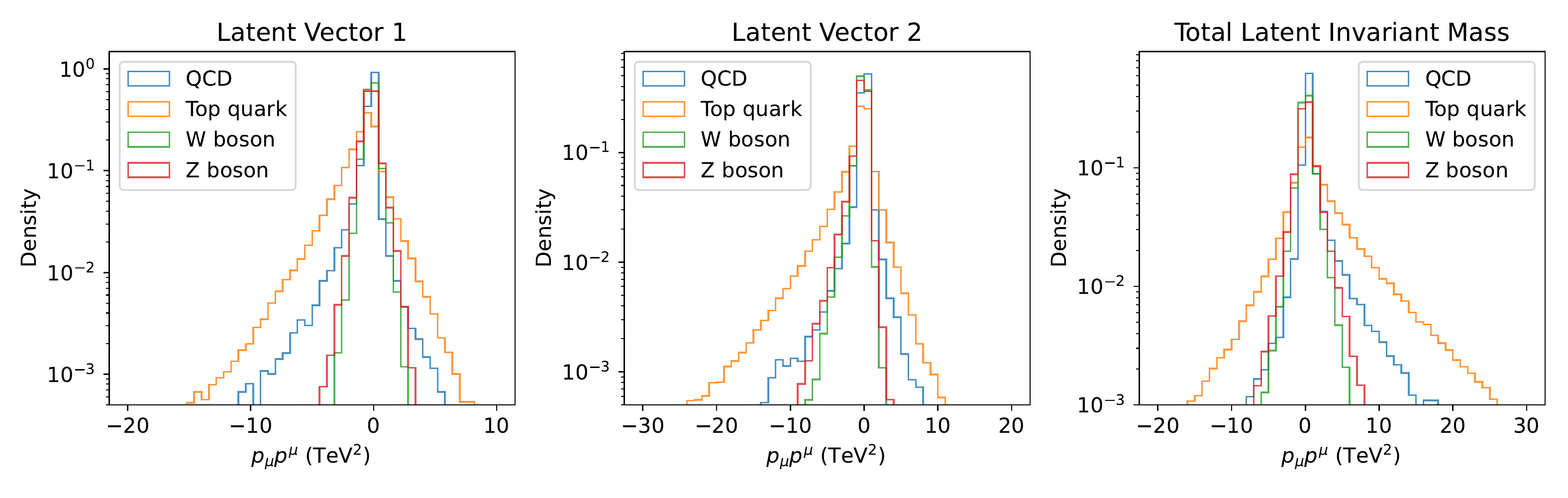}
    \includegraphics[width=\linewidth]{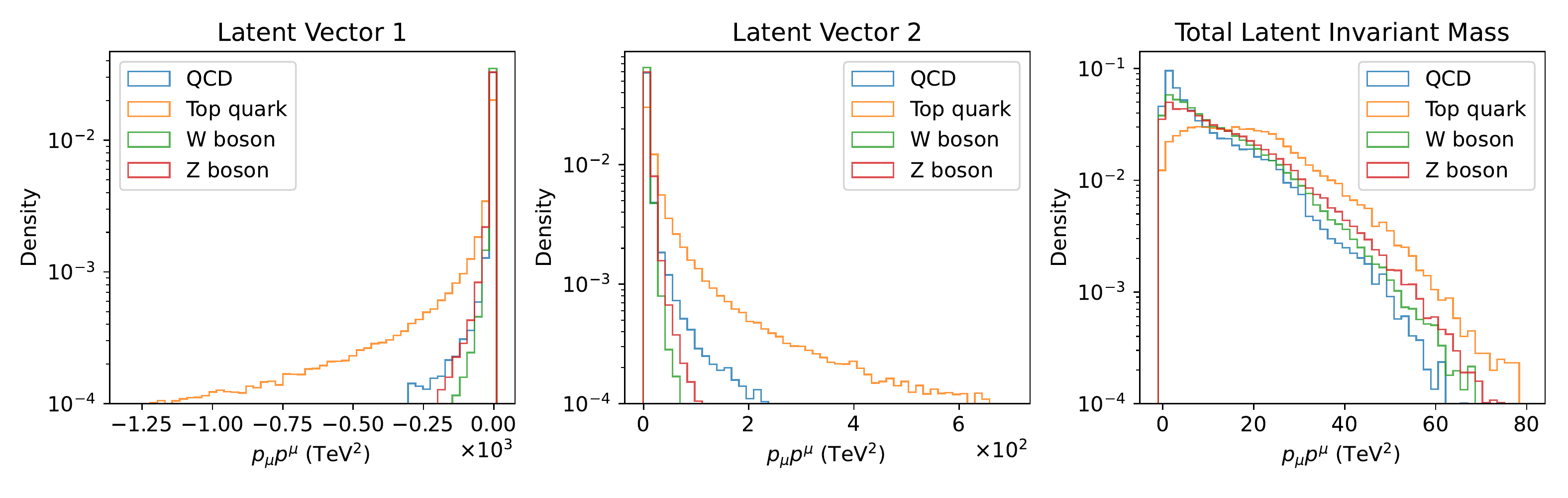}
    \caption{
        \textbf{Top}: distributions of the invariant mass squared of the latent 4-vectors and jet momenta of the LGAE-Mix with $\tau_{(1/2, 1/2)} = 2$ latent 4-vectors.
        \textbf{Bottom}: distributions of the invariant mass squared of two latent 4-vectors and jet momenta of the LGAE-Min-Max with $\tau_{(1/2, 1/2)} = 2$ latent 4-vectors.
    }
    \label{fig:distribution}
\end{figure*}

The outputs of the LGAE encoder are irreducible representations of the Lorentz groups; they consist of a pre-specified number of Lorentz scalars, vectors, and potentially higher-order representations.
This implies a significantly more interpretable latent representation of the jets than traditional autoencoders, as the information distributed across the latent space is now disentangled between the different irreps of the Lorentz group. 
For example, scalar quantities like the jet mass will necessarily be encoded in the scalars of the latent space, and jet and particle 4-momenta in the vectors. 

We demonstrate the latter empirically on the LGAE-Mix model ($\tau_{(1/2, 1/2)} = 2$) by looking at correlations between jet 4-momenta and the components of different combinations of latent vector components.
Figure~\ref{fig:correlations} shows that, in fact, the jet momenta is encoded in the imaginary component of the sum of the latent vecotrs.

We can also attempt to understand the anomaly detection performance by looking at the encodings of the training data compared to the anomalous signal.
Figure~\ref{fig:distribution} shows the individual and total invariant mass of the latent vectors of sample LGAE models for QCD and top quark, W boson, and Z boson inputs.
We observe that despite the overall similar kinematic properties of the different jet classes, the distributions for the QCD background are significantly different from the signals, indicating that the LGAE learns and encodes the difference in jet substructure --- despite substructure observables such as jet mass not being direct inputs to the network --- explaining the high performance in anomaly detection.

Finally, while in this section we showcased simple ``brute-force'' techniques for interpretability by looking directly at the distributions and correlations of latent features, we hypothesize that such an equivariant latent space would also lend itself effectively to the vast array of existing explainable AI algorithms~\cite{vilone_xaireview,minh_xaireview}, which generically evaluate the contribution of different input and intermediate neuron features to network outputs.
We leave a detailed study of this to future work.

\subsection{Data efficiency}
\label{sec:data-efficiency}
\begin{figure}
    \centering
    \includegraphics[width=\linewidth]{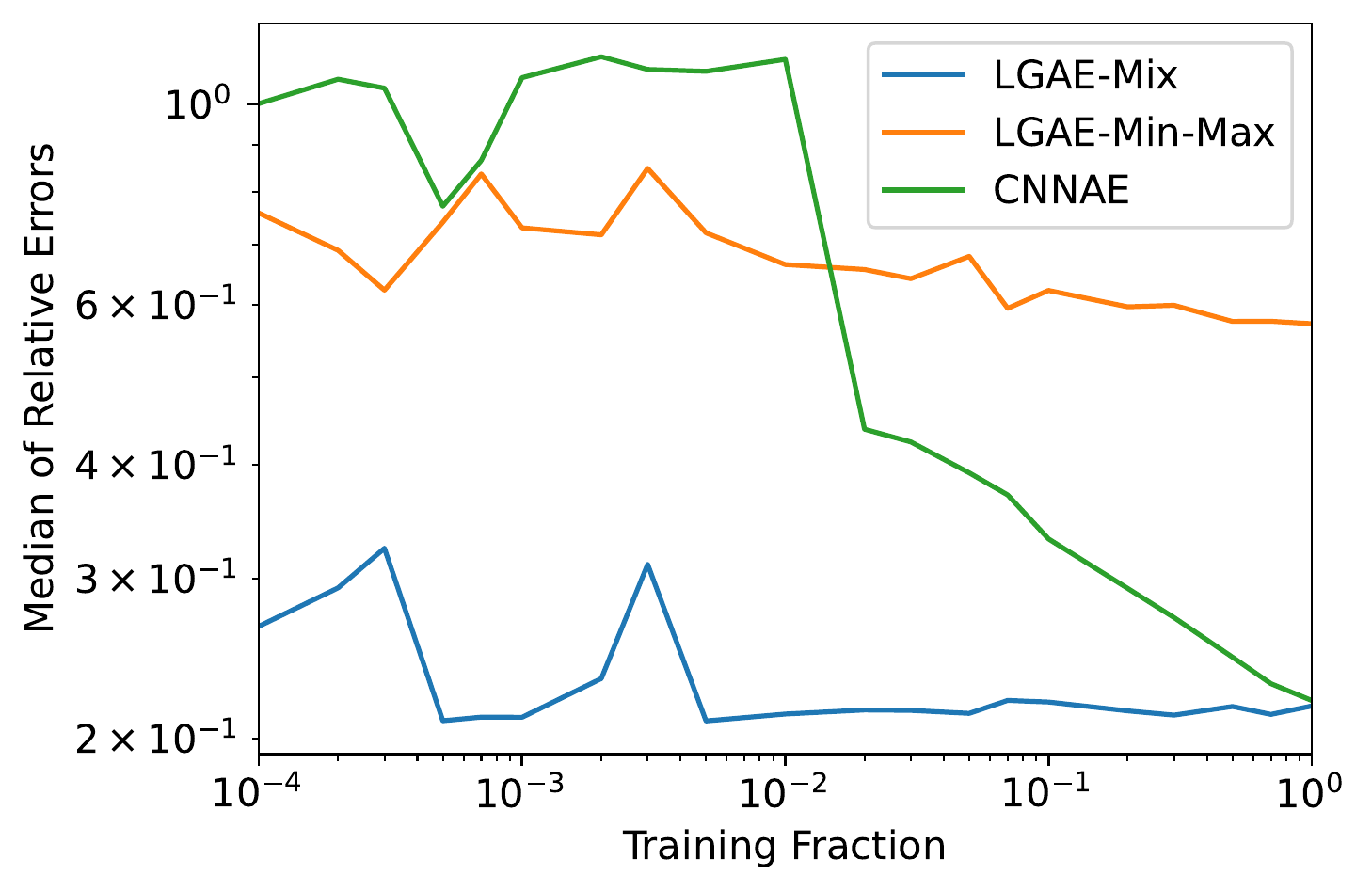}
    \caption{Median magnitude of relative errors of jet mass reconstruction by LGAE and CNNAE models at trained on different fractions of the training data.}
    \label{fig:median-jet-mass}
\end{figure}


In principle, equivariant neural networks should require less training data for high performance, since critical biases of the data, which would otherwise have to be learnt by non-equivariant networks, are already built in.
We test this claim by measuring the performances of the best-performing LGAE and CNNAE architectures from Sec.~\ref{sec:reconstruction} trained on varying fractions of the training data.

The median magnitude of the relative errors between the reconstructed and true jet masses of the different models and fractions is shown in Fig.~\ref{fig:median-jet-mass}.
Each model is trained five times per training fraction, with different random seeds, and evaluated on the same-sized validation dataset; the median of the five models is plotted.
We observe that, in agreement with our hypothesis, the LGAE models both maintain their high performance all the way down to training on 1\% of the data, while the CNNAE's performance steadily degrades down to 2\% and then experiences a further sharp drop.


\section{Conclusion}
\label{sec:conclusion}
We develop the Lorentz group autoencoder (LGAE), an autoencoder model equivariant to Lorentz transformations.
We argue that incorporating this key inductive bias of high energy physics (HEP) data can have a significant impact on the performance, efficiency, and interpretability of machine learning models in HEP.
We apply the LGAE to tasks of compression and reconstruction of input quantum chromodynamics (QCD) jets, and of identifying anomalous top quark, W boson, and Z boson jets. 
We report excellent performance in comparison to baseline graph and convolutional neural network autoencoder models, with the LGAE outperforming them on several key metrics.
We also demonstrate the LGAE's interpretability, by analyzing the latent spaces of LGAE models for both tasks, and data efficiency relative to baseline models.
The LGAE opens many promising avenues in terms of both performance and model interpretability, with the exploration of new datasets, the magnitude of Lorentz and permutation symmetry breaking due to detector effects, higher-order Lorentz group representations, and challenges with real-life compression and anomaly detection applications all exciting possibilities for future work.

\begin{acknowledgements}
    We would like to thank Dr. Rose Yu for discussions on equivariant neural networks, and Dr. Dylan Rankin for suggestions on the anomaly detection performance and latent space analysis of the LGAE.
    Z.~H. thanks the UC San Diego Faculty Mentor Program for supporting this research.
    R.~K. was partially supported by the LHC Physics Center at Fermi National Accelerator Laboratory, managed and operated by Fermi Research Alliance, LLC under Contract No. DE-AC02-07CH11359 with the U.S. Department of Energy (DOE).
    J.~D. is supported by the DOE, Office of Science, Office of High Energy Physics Early Career Research program under Award No. DE-SC0021187, the DOE, Office of Advanced Scientific Computing Research under Award No. DE-SC0021396 (FAIR4HEP), and the NSF HDR Institute for Accelerating AI Algorithms for Data Driven Discovery (A3D3) under Cooperative Agreement OAC-2117997.
    N.~C. was supported by the European Research Council (ERC) under the European Union's Horizon 2020 research and innovation program (Grant Agreement No. 772369).
    This work was performed using the Pacific Research Platform Nautilus HyperCluster supported by NSF awards CNS-1730158, ACI-1540112, ACI-1541349, OAC-1826967, the University of California Office of the President, and the University of California San Diego's California Institute for Telecommunications and Information Technology/Qualcomm Institute. 
    Thanks to CENIC for the 100\,Gpbs networks.
\end{acknowledgements}

\vspace{0.1mm}

\small{
\noindent
\textbf{Data Availability Statement}
The datasets used in this manuscript are publicly available on Zenodo~\cite{jetnetdataset} and the code for all models used in this paper can be found in a public repository~\cite{LGAE_code}.
}

\appendix
\section{Model details}
\label{app:hyperparams}

\subsection{LGAE}

For both encoder and decoder, we choose $N_\mathrm{MP}^\mathrm{E} = N_\mathrm{MP}^\mathrm{D} = 4$ LMP layers.
The multiplicity per node in each LMP layer has been optimized to be
\begin{equation}
    \left\{ \parenthesis{\tau_{(m, n)}^{(t)}}^\mathrm{E} \right\}_{t=1}^{4} = (3,3,4,4)
\end{equation}
for the encoder and
\begin{equation}
    \left\{ \parenthesis{\tau_{(m, n)}^{(t)}}^\mathrm{D} \right\}_{t=1}^{4} = (4,4,3,3)
\end{equation}
for the decoder, the components in the vector on the right-hand side are the multiplicity in each of the four LMP layers per network, and the multiplicity per layer is the same for all representations.
After each CG decomposition, we truncate irreps of dimensions higher than $(1/2, 1/2)$ for tractable computations, i.e., after each LMP operation we are left with only scalar and vector representations per node.
Empirically, we did not find such a truncation to affect the performance of the model.
This means that the LMP layers in the LGAE are similar in practice to those of LorentzNet, which uses only scalar and vector representations throughout, but are more general as higher dimensional representations are involved in the intermediate steps before truncation.

The differentiable mapping $f(d_{ij})$ in Eq.(\ref{eq:msg}) is chosen to be the Lorentzian bell function as in Ref.~\cite{LGN}.
For all models, the latent space contains only $\tau_{(0,0)} = 1$ complex Lorentz scalar, as we found increasing the number of scalars beyond one did not improve the performance in either reconstruction or anomaly detection.
Empirically, the reconstruction performance increased with more latent vectors, as one might expect, while anomaly detection performance generally worsened from adding more than two latent vectors.

\subsection{GNNAE}
\begin{figure}[htpb]
    \centering
    \includegraphics[scale=0.55]{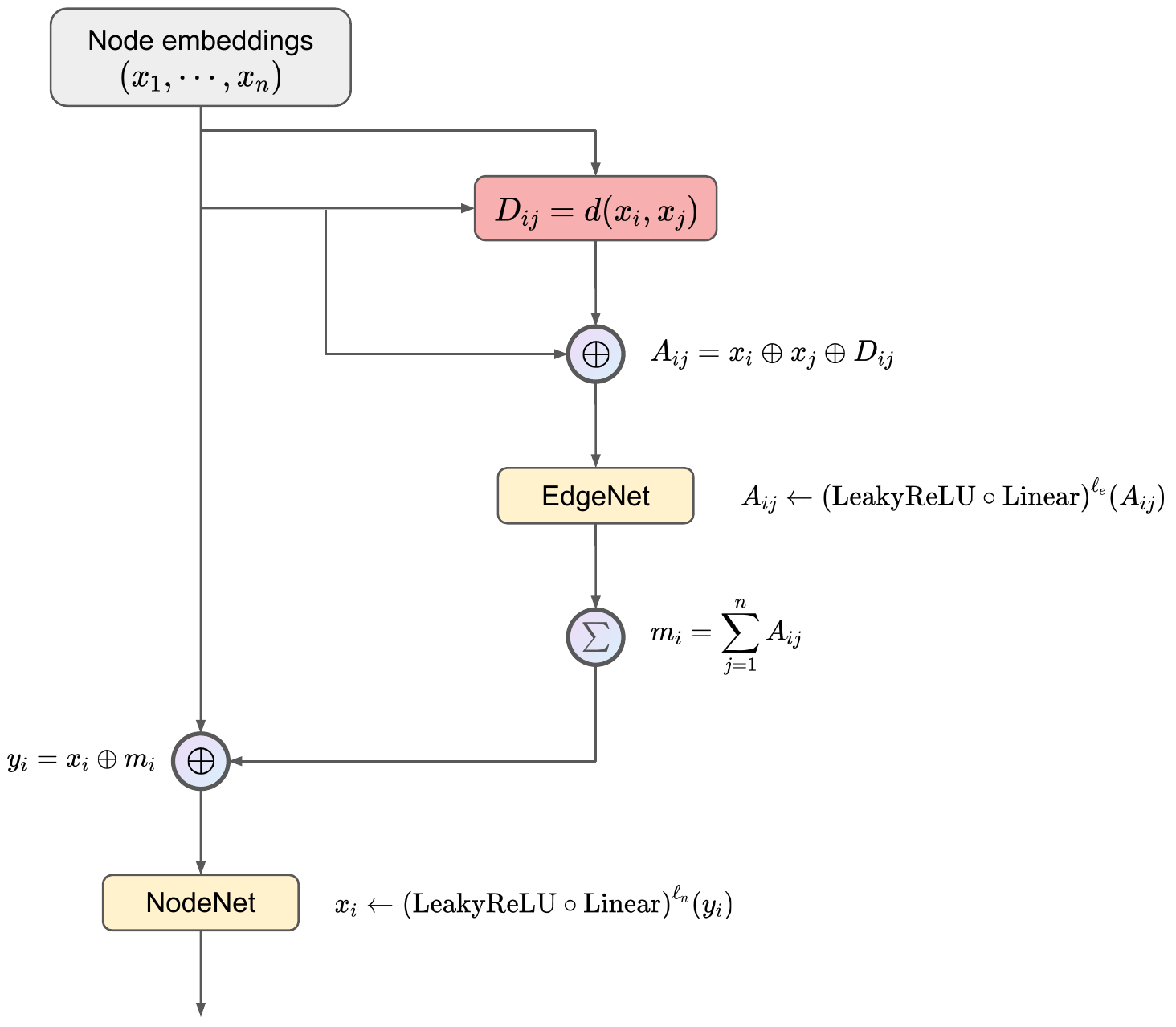}
    \caption{An MPNN layer in the GNNAE. Here, $\mathrm{EdgeNet}$ and $\mathrm{NodeNet}$ are feed-forward neural networks.}
    \label{fig:gnn-message-passing}
\end{figure}

The GNNAE is constructed from fully-connected MPNNs.
The update rule in the $(t+1)$-th MPNN layer is based on Ref.~\cite{Raghav_GAN}, and given by
\begin{align}
    m_i^{(t)}   & = \sum_{j=1}^n f_e^{(t)}\left(
    x_i^{(t)} \oplus x_j^{(t)} \oplus d\left(x_i^{(t)}, x_j^{(t)}\right)
    \right), \label{eq:gnnae-m}                  \\
    x_i^{(t+1)} & = f_n^{(t)} \left(
    x_i^{(t)} \oplus m_i^{(t)}
    \right), \label{eq:gnnae-x}
\end{align}
where $x_i^{(t)}$ is the node embedding of node $i$ at $t$-th iteration,
$d$ is any distance function (Euclidean norm in our case),
$m_i^{(t)}$ is the message for updating node embedding in node $i$,
$f_e^{(t+1)}$ and $f_n^{(t+1)}$ are any learnable mapping at the current MP layer.
A diagram for an MPNN layer is shown in Fig.~\ref{fig:gnn-message-passing}.
The overall architecture is similar to that in Fig.~\ref{fig:architecture-full}, with the LMP replaced by the MPNN.
The code for the GNNAE model can be found in the Ref.~\cite{GNNAE_code}.

For both the encoder and decoder, there are $3$ MPNN layers.
The learnable functions in each layer are optimized to be
\begin{equation}
    \begin{split}
        f_n^{(1)} &= (
        \mathrm{LeakyReLU}_{0.2}
        \circ
        \mathrm{Linear}_{30\to 15}) \\
        &\quad \circ
        (\mathrm{LeakyReLU}_{0.2}
        \circ
        \mathrm{Linear}_{60\to 30}) \\
        f_e^{(1)} &=
        (\mathrm{LeakyReLU}_{0.2}
        \circ
        \mathrm{Linear}_{40\to 30}), \\
        &\quad \circ
        (\mathrm{LeakyReLU}_{0.2}
        \circ
        \mathrm{Linear}_{50\to 40}) \\
        &\quad \circ
        (\mathrm{LeakyReLU}_{0.2}
        \circ
        \mathrm{Linear}_{61\to 50}),
    \end{split}
\end{equation}
\begin{equation}
    \begin{split}
        f_n^{(2)} &= (
        \mathrm{LeakyReLU}_{0.2}
        \circ
        \mathrm{Linear}_{15\to 8}) \\
        &\quad \circ
        (\mathrm{LeakyReLU}_{0.2}
        \circ
        \mathrm{Linear}_{45\to 15}) \\
        f_e^{(2)} &=
        (\mathrm{LeakyReLU}_{0.2}
        \circ
        \mathrm{Linear}_{31\to 30}), \\
        &\quad \circ
        (\mathrm{LeakyReLU}_{0.2}
        \circ
        \mathrm{Linear}_{30\to 30}) \\
        &\quad \circ
        (\mathrm{LeakyReLU}_{0.2}
        \circ
        \mathrm{Linear}_{30\to 30}),
    \end{split}
\end{equation}
\begin{equation}
    \begin{split}
        f_n^{(3)} &= (
        \mathrm{LeakyReLU}_{0.2}
        \circ
        \mathrm{Linear}_{8 \to \delta}) \\
        &\quad \circ
        (\mathrm{LeakyReLU}_{0.2}
        \circ
        \mathrm{Linear}_{38\to 8}) \\
        f_e^{(3)} &=
        (\mathrm{LeakyReLU}_{0.2}
        \circ
        \mathrm{Linear}_{20\to 30}), \\
        &\quad \circ
        (\mathrm{LeakyReLU}_{0.2}
        \circ
        \mathrm{Linear}_{16\to 20}) \\
        &\quad \circ
        (\mathrm{LeakyReLU}_{0.2}
        \circ
        \mathrm{Linear}_{17\to 16}),
    \end{split}
\end{equation}
where $\mathrm{LeakyRelu}_{0.2}(x) = \max(0.2 x, x)$ is the LeakyReLu function.

Depending on the aggregation layer, the value of $\delta$ in $f_n^{(3)}$ and the final aggregation layer is different.
For GNNAE-JL encoders, $\delta = N \times \dim(L)$, where $L$ is the latent space, and $N$ is the number of nodes in the graph.
Then, mean aggregation is done across the graph.
For GNNAE-PL encoders, $\delta = d$, where $d$ is the node dimension in the latent space.
In the GNNAE-JL decoder, the input layer is a linear layer that recovers the particle cloud structure similar to that in the LGAE.

\subsection{CNNAE}
The encoder is composed of two convolutional layers with kernel size $(3, 3)$, stride size $(2, 2)$, ``same" padding, and $128$ output channels, each followed by a ReLU activation function. 
The aggregation layer into the latent space is a fully-connected linear layer. 
The decoder is composed of transposed convolution layers (also known as deconvolutional layers) with the same settings as the encoder. 
A softmax function is applied at the end so that the sum of all pixel values in an image is $1$, as a property of the jet image representation.
A 55-dimensional latent space  is chosen so that the compression rate is $55/90 \approx 60\%$ for even comparisons with the LGAE and GNNAE models.


\section{Training details}
\label{app:training}
We use the Chamfer loss function~\cite{10.5555/1622943.1622971,Fan_2017_CVPR,Zhang2020FSPool} for the LGAE-Min-Max and GNNAE-JL models, and  MSE for LGAE-Mix and GNNAE-PL.
We tested the Hungarian loss~\cite{hungarian,scipy} and differentiable energy mover's distance (EMD)~\cite{emd}, calculated using the \jetnet library~\cite{jetnetlibrary}, as well but found the Chamfer and MSE losses more performant.

The graph-based models are optimized using the Adam optimizer~\cite{ADAM}
implemented in \textsc{PyTorch}~\cite{pytorch}
with a learning rate $\gamma = 10^{-3}$, coefficients $(\beta_1, \beta_2) = (0.9, 0.999)$, and weight decay $\lambda = 0$.
The CNNAE is optimized using the same optimizer implemented in TensorFlow~\cite{tensorflow}. 
They are all trained on single NVIDIA RTX 2080 Ti GPUs each for a maximum of 20000 epochs using early stopping with the patience of 200 epochs.
The total training time for LGAE models is typically 35 hours, and at most 100 hours, while GNNAE-PL and GNNAE-JL train for 50 and 120 hours on average, respectively.
By contrast, the CNNAE model, due to its simplicity, can typically converge within 3 hours.

\section{Equivariance tests}
\label{app:equivariancetests}

We test the covariance of the LGAE models to Lorentz transformations and find they are indeed equivariant up to numerical errors.
Bogatskiy \etal point out that equivariance to boosts in particular is sensitive to numerical precision~\cite{LGN}, so we use double precision (64-bit) throughout the model. 
In addition, we scale down the data by a factor of 1,000 (i.e. working in the units of PeV) for better numerical precision at high boosts.

\begin{figure}[ht]
    \centering
    \includegraphics[width=\linewidth]{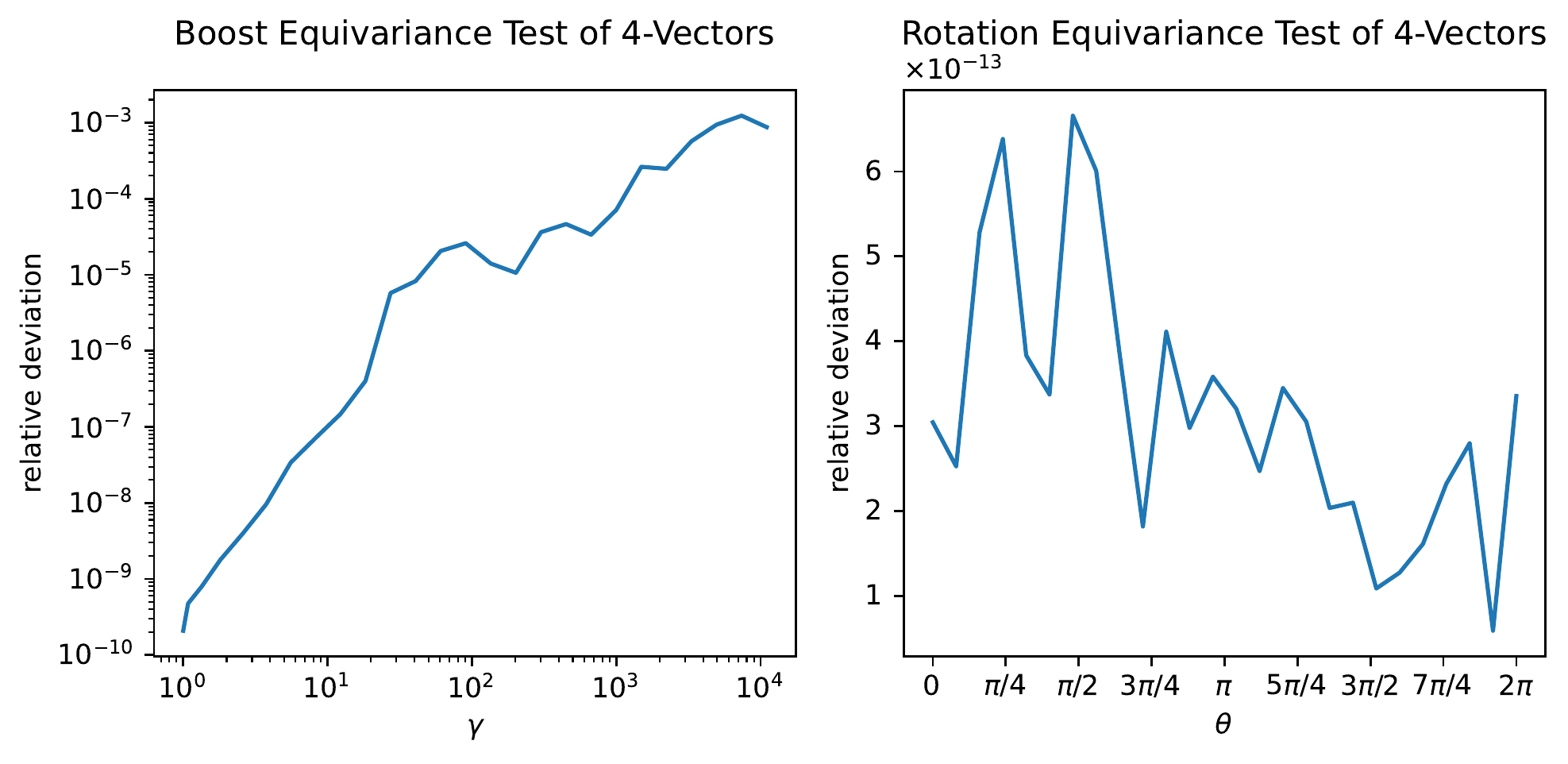}
    \caption{The relative deviations, as defined in Eq.~(\ref{eq:delta-p}), of the output 4-momenta $p^\mu$ to boosts along the $z$-axis (left) and rotations around the $z$-axis (right).}
    \label{fig:equivariance}
\end{figure}

For a given transformation $\Lambda \in \mathrm{SO}^+(3,1)$ we compare $\Lambda \cdot \mathrm{LGAE}(p)$ and $ \mathrm{LGAE}(\Lambda \cdot p)$ are compared, where $p$ is the particle-level 4-momentum. 
The relative deviation is defined as
\begin{equation} \label{eq:delta-p}
    \delta_p(\Lambda) = \left|\frac{
        \mathrm{mean}(\mathrm{LGAE}(\Lambda \cdot p)) -
        \mathrm{mean}(\Lambda \cdot \mathrm{LGAE}(p))
    }{\mathrm{mean}(\Lambda \cdot \mathrm{LGAE}(p))}\right|
\end{equation}
Figure~\ref{fig:equivariance} shows the mean relative deviation, averaged over each particle in each jet, over $3000$ jets from our test dataset from boosts along and rotations around the $z$-axis.
We find the relative deviation from boosts to be within $\order{10^{-3}}$ in the interval $\gamma \in [0, \cosh(10)]$ (equivalent to $\beta \in [0, 1 - 4\times10^{-9}$]) and from rotations to be $<10^{12}$. 


\bibliographystyle{apsrev4-1}       
\bibliography{bibliography.bib}

\end{document}